# Generation of very flat optical frequency combs from continuous-wave lasers using cascaded intensity and phase modulators driven by tailored radio frequency waveforms

Rui Wu, V. R. Supradeepa,[*] Christopher M. Long, Daniel E. Leaird and Andrew M. Weiner

*School of Electrical and Computer Engineering, Purdue University, 465 Northwestern Avenue,*

*West Lafayette, Indiana 47907, USA*

[*]Corresponding author: supradeepa@purdue.edu

We demonstrate a scheme based on a cascade of lithium niobate intensity and phase modulators driven by specially tailored radio frequency waveforms to generate an optical frequency comb with very high spectral flatness. In this work we demonstrate a 10 GHz comb with 38 comb lines within a spectral power variation below 1-dB. The number of comb lines that can be generated is limited by the power handling capability of the phase modulator, and this can be scaled without compromising the spectral flatness. Furthermore, the spectral phase of the generated combs in our scheme is almost purely quadratic which, as we will demonstrate, allows for high quality pulse compression using only single mode fiber.

OCIS codes: (060.0060) Fiber optics and optical communications; (060.5060) Phase modulation; (060.5625) Radio frequency photonics; (120.3940) Metrology; (320.5540) Pulse shaping; (320.5520) Pulse compression;

Strong sinusoidal phase modulation of a continuous wave (CW) laser creates multiple sidebands leading to generation of a frequency comb [1]. Advantages of this technique include the ability to create high repetition rate combs with stable optical center frequencies given by the source laser and convenient tuning of the repetition rate and optical center frequency. Therefore, such combs are a source of choice for some applications in optical communications [2], Radio frequency (RF) photonics [3] and optical arbitrary waveform generation (OAWG) [4]. However, by phase modulation alone, the spectral flatness of the comb is quite poor with significant line to line amplitude variations.

A flat frequency comb is generally desired. For example, if the comb is used as a multi-wavelength source, it is desirable to have equal power in different wavelengths. For applications such as RF photonic filtering requiring a specific spectral shape ([5]), apodization is much easier with minimal excess loss if we start off with a flat comb. Also, for time domain applications where a short pulse is needed, abrupt comb line to line variations result in poor pulse quality. Some examples of recent research activity based on modulation of a CW laser aimed at generating flatter combs can be found in [6-10]. Though all of these schemes improve the flatness significantly compared to phase-only modulation, they still provide either limited flatness over the bandwidth of interest or limited number of comb lines over which flatness can be maintained. For example, in [7] the comb shows 7-dB power variation over the ~60 comb lines which constitute the flat region of the spectrum, while references [8] and [9] report 11 comb lines, with 1.1dB and 1.9dB power variation, respectively. In this work, we demonstrate a scheme to achieve very flat combs using a cascade of intensity and phase modulators driven by tailored RF waveforms. In particular, we report a 10 GHz comb with 38 comb lines in a 1-dB bandwidth and ~60 comb lines above the noise floor of the OSA (60-dB bandwidth). The number of comb lines can be scaled by increasing the RF drive power without sacrificing spectral flatness. Also, the spectral phase of the comb generated in our experiment is almost purely quadratic, which allows for generation of high quality pulses via compression in a simple dispersive fiber or chirped fiber Bragg grating.

Our research draws on the interesting theoretical work of V. Torres-Company and coworkers [11], which proposed a novel way of understanding the spectral flatness achieved in previous schemes such as [6] and [7]. When a flat topped pulse is subjected to a strong, periodic, quadratically varying temporal phase, it undergoes time-to-frequency mapping [12, 13], resulting in a flat comb due to the shape being similar to the time domain intensity of the input waveform. A convenient approach to generate flat topped pulses is to use an intensity modulator driven with a sinusoid with an amplitude $V = V_\pi / 2$ of the modulator and a DC bias corresponding to a phase shift of $\phi_{dc} = -\pi/2$ (the envelope of the output field after the intensity modulator is given by the relation $(1 + \exp(j\phi_{dc})\exp(j\pi V/V_\pi \cos\omega_{rf} t))$ which in our case, after substitution simplifies to $(1 + \exp(-j\frac{\pi}{2})\exp(j\frac{\pi}{2}\cos\omega_{rf} t)))$. Generating a periodic quadratic temporal phase though is hard; however, a sinusoidal temporal phase can be approximated by a quadratic function around its peak or its valley and to a first order provides a means to generate the required quadratic temporal phase. Fig 1(a) shows this scheme combining the two aspects, and Fig 1(b) shows the spectrum simulated assuming the phase modulator is driven with a sinusoidal voltage whose amplitude $V$ satisfies the relation $\pi V/V_\pi = 20$ (i.e. the effect of the phase modulator can be written as a multiplication by $\exp(j\pi V/V_\pi \cos\omega_{rf} t) = \exp(j20\cos\omega_{rf} t)$). In Fig. 1(b) we see a flat central section to the spectra, but towards the edges there are pronounced "bat ears." To understand this, we look at figure 1(c) (blue, solid) which shows the time domain signal, and 1(c) (red, solid) which shows the sinusoidal temporal phase. In the central part, the time domain signal is very flat and the phase is almost quadratic leading to good spectral flatness. As we move away from the center and the phase starts deviating from a quadratic, the flatness is degraded. The strong peaks or bat-ears occur at the spectral extrema, where the instantaneous frequency, i.e., the time derivative of the temporal phase, is at a minimum or maximum. The bat-ears may be explained on the basis of the relatively long time for which the instantaneous frequency function dwells at the frequency extrema [14].

In order to obtain flat spectra, we need both a nice flat-topped pulse as well as a good quadratic temporal phase. Accordingly, we propose two modifications to the above method. A first improvement would be to make the flat-topped pulse sharper, which will reduce the fraction of the waveform seeing a significant departure from quadratic temporal phase. This can be achieved using two intensity modulators in series driven with the same parameters as described previously. Figure 1(c) (black, dashed) shows the new sharper time domain waveform. Secondly, since the bat ears happen due to deviation from quadratic phase, we attempt to generate a phase profile that better approximates a quadratic. The expansion for the cosine around its maximum up to the first undesirable term is,

$$\cos(\omega_{rf} t) = 1 - \frac{(\omega_{rf} t)^2}{2!} + \frac{(\omega_{rf} t)^4}{4!} - \ldots \tag{1}$$

So we have

$$\cos(\omega_{rf} t) - \frac{1}{16}\cos(2\omega_{rf} t) = \frac{15}{16} - 0.75\frac{(\omega_{rf} t)^2}{2!} + 0 + \ldots \tag{2}$$

From eqn (2) we see that by using the first and $2^{nd}$ harmonics (which can be easily generated with a X2 frequency multiplier) with a suitable ratio and phase shift between them, the $4^{th}$ order term (which is the first non-ideal term) can be made zero. With regard to higher order terms, since the cascade of the intensity modulators creates a waveform with a sharper decay, they are less of an issue in the region of the flat topped pulse. Figure 2(a) shows the scheme incorporating both the improvements we discussed. A tunable attenuator adjusts the ratio between the fundamental harmonic and the 2$^{nd}$ harmonic to the required value ((1/16)^2 ~ -24dB) and a tunable RF phase shifter ensures that they are 180degrees out of phase as required. We also have other tunable RF phase shifters to ensure that there is timing match between the different components. Figure 2(b) shows the simulated spectrum, which is now significantly flattened, with the absence of any bat-ear shape. In our actual experiment, a minor difference to the scheme shown in fig 2(a) is that we use two phase modulators in series instead of one. This is a common technique used (e.g., [7]) to overcome the RF power handling limit of the phase modulators. In our case we drive each phase modulator by its maximum RF power of 1W; driving two phase modulators identically allows greater power handling and doubles the number of comb lines. The IM we use has a $V_\pi$ of ~8V, the PM has a $V_\pi$ of ~3V and all modulators are pigtailed with polarization maintaining fiber. The frequency of the RF oscillator is 10GHz. The net optical loss of the whole apparatus is ~15dB (the unavoidable loss due to gating of the CW by the intensity modulators is only ~4dB, while the rest is mainly modulator insertion losses). Figures 2(c) and 2(d) show the experimental output comb spectra in linear and log scale, respectively. We observe a very flat spectral profile with 38 comb lines in a 1-dB bandwidth ($<\pm$ 10%) (circled in fig 2(d)) with much smaller variations in smaller bandwidths around the center.

Another interesting aspect is that, in the large phase modulation limit, a waveform with a pure quadratic temporal phase corresponds to a spectrum with pure quadratic phase. This in turn corresponds to a pulse with a linear chirp, which can be compressed to the bandwidth limit with high quality just using an appropriate length of standard single mode fiber (compared to more complex pulse shaper based methods [15]). For the bandwidths involved, the effect of dispersion slope of the fiber is small. Figure 3(a) (blue, solid) which shows the simulated spectral phase of the comb together with a quadratic fit (red, dashed) clearly indicates this by showing an excellent agreement between them. This idea was first discussed in [16] and the process of making the drive more quadratic is referred to as aberration correction. To demonstrate it's superiority over a pure sinusoid, fig 3(b) shows the simulated spectral phase for the same configuration with the phase modulator now driven by a pure sinusoid. In this case, the spectral phase of the comb differs more significantly from a quadratic; this is expected to translate into lower quality pulse compression.

We measured the spectral phase of the comb using our frequency comb characterization technique based on a linear implementation of spectral shearing interferometry, described in detail in [17]. The measured phase indicated that pulse compression could be accomplished using ~850m of SMF. Fig 3(c) and 3(d) (blue, solid) show the measured intensity autocorrelation of the pulse resulting after fiber propagation in a short and wide temporal window. The theoretical intensity autocorrelation (red, dashed) taking into account the measured comb spectrum and assuming flat phase is also plotted. We see excellent agreement, indicating high quality pulse compression to a bandwidth-limited pulse. The obtained pulse has an intensity autocorrelation full width at half maximum (FWHM) of ~2.8ps, which corresponds to an intensity FWHM of ~2.1ps assuming the sinc^2 intensity profile appropriate for the nearly flat-topped spectrum.

In summary, we have demonstrated a new, easily scalable scheme for generating very flat optical frequency combs using cascaded intensity and phase modulators driven by tailored

RF waveforms. In this work we demonstrated a 10 GHz comb with 38 comb lines in a 1-dB bandwidth and around 60 comb lines in total. Another attractive aspect of our scheme is the ability to achieve very high quality compression (in our experiments resulting in a ~2.1 ps bandwidth-limited pulse) simply through propagation in standard single-mode fiber.

**Acknowledgements**

This work was supported in part by the National Science Foundation under grant ECCS-0601692 and by the Naval Postgraduate School under grant N00244-09-1-0068 under the National Security Science and Engineering Faculty Fellowship program.

Fig.1 (a) Experimental scheme for flattening frequency combs generated by phase modulation using cascaded intensity modulator (IM) and phase modulator (PM), PS – RF phase shifter (b) Simulated output spectrum for this scheme, (c) Time domain plot showing the temporal phase applied by the PM (red, solid) to the outputs after 1 IM (blue, solid) and 2 IMs (black, dashed)

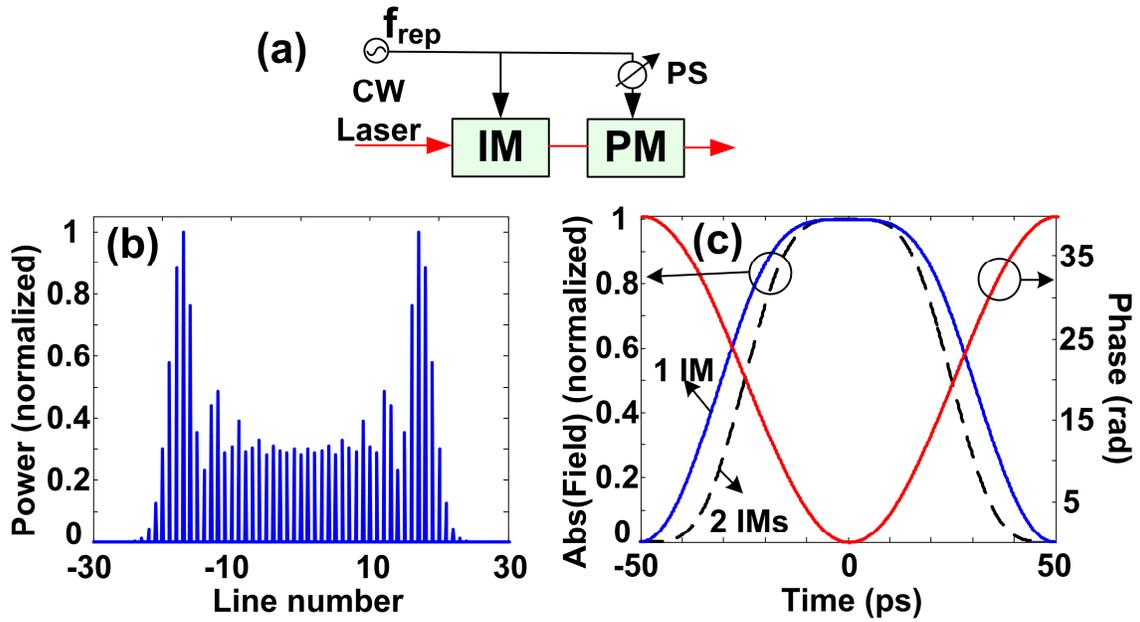

Fig.2 (a) Experimental scheme for comb generation with high spectral flatness, VA – variable attenuator, PS – RF phase shifters, (in our actual experiment, we use two phase modulators in series to generate greater number of lines) (b) Simulated output spectrum in this case, (c), (d) Output spectra of the experimentally generated comb in linear and log scale showing 38 comb lines within 1-dB variation

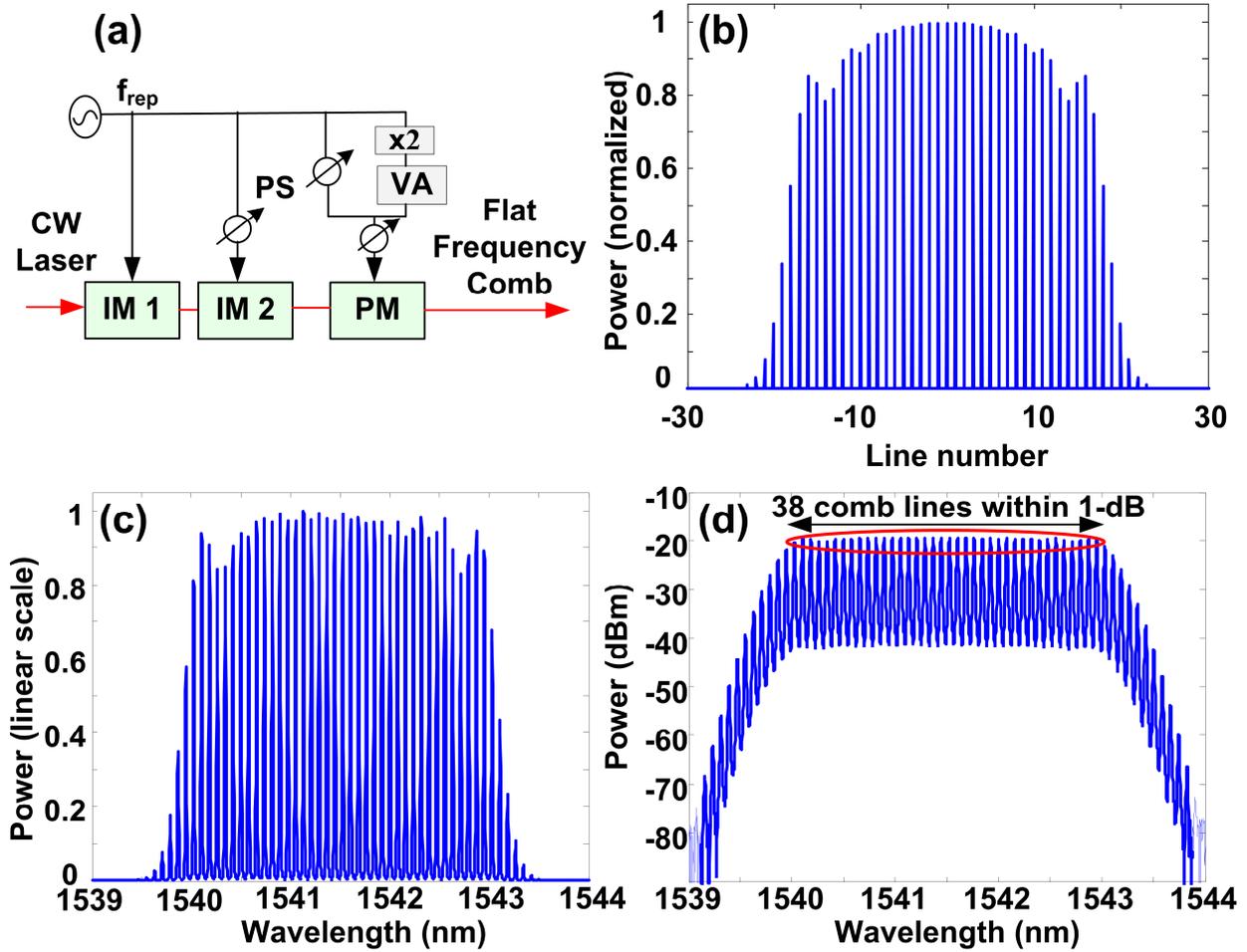

Fig.3, (a), (b) Simulated spectral phase and the quadratic fit to it for the case of phase modulation with tailored RF waveform (a) and a sinusoid (b). (c), (d) Short and long aperture time domain intensity autocorrelations of the output pulse (measured, blue, solid) (obtained after comb propagation through ~850m of SMF) superimposed with the autocorrelation calculated taking the spectra (2(c)) and assuming flat spectral phase. A very good agreement is seen between the two.

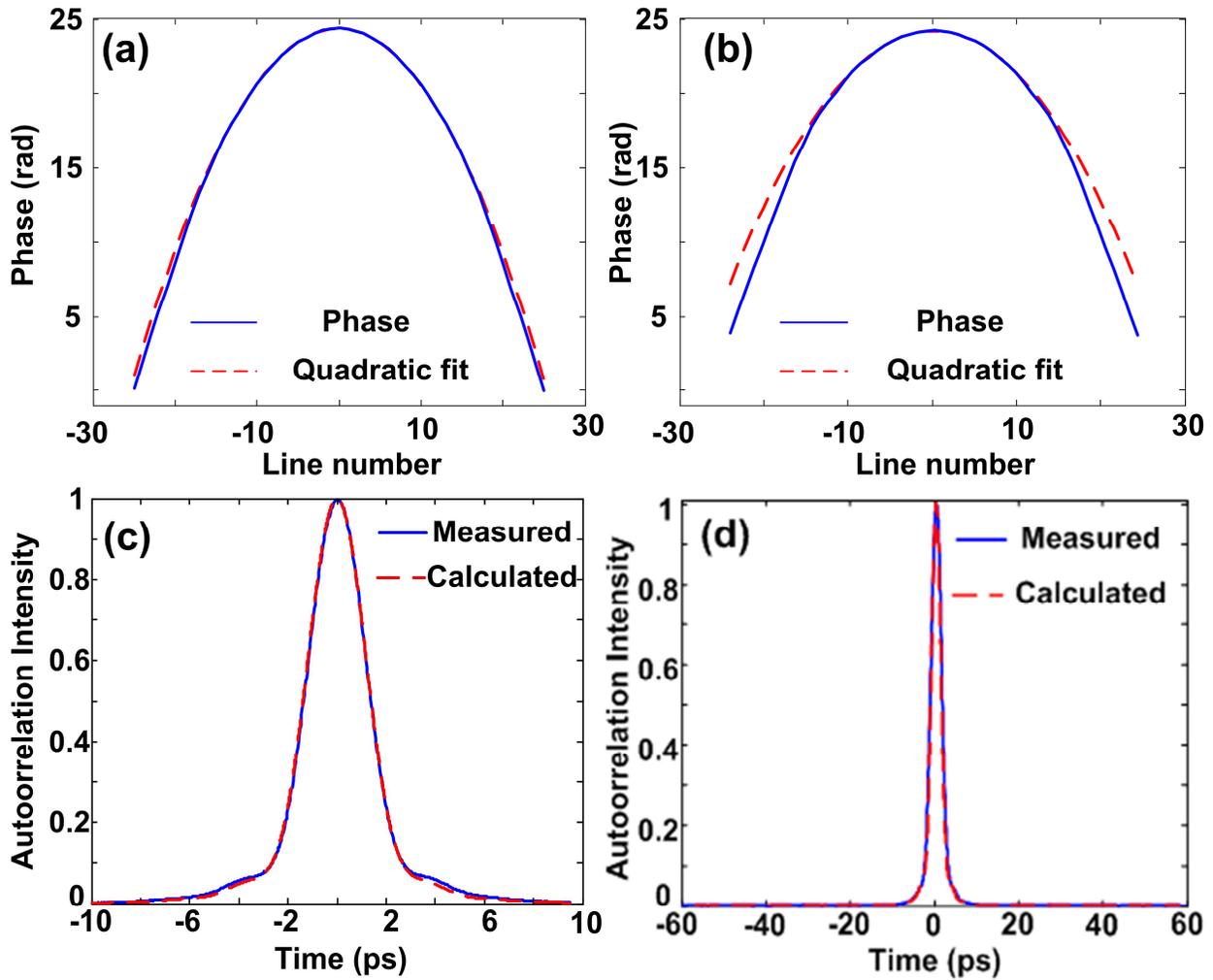